\documentclass[twocolumn]{aastex631}

\begin{document}

\newcommand{\kms}{km~s$^{-1}$}	\newcommand{\cms}{cm~s$^{-2}$}
\newcommand{\msun}{$M_{\odot}$} \newcommand{\rsun}{$R_{\odot}$} 
\newcommand{\teff}{$T_{\rm eff}$} \newcommand{\logg}{$\log{g}$} 
\newcommand{\mas}{mas~yr$^{-1}$}

\title{ The ELM Survey. IX.  A Complete Sample of Low Mass White Dwarf Binaries in 
the SDSS Footprint }

\author[0000-0002-4462-2341]{Warren R.\ Brown} \affiliation{Center for 
Astrophysics, Harvard \& Smithsonian, 60 Garden Street, Cambridge, MA 02138 USA}

\author[0000-0001-6098-2235]{Mukremin Kilic} \affiliation{Homer L. Dodge Department 
of Physics and Astronomy, University of Oklahoma, 440 W. Brooks St., Norman, OK, 
73019 USA}

\author[0000-0002-9878-1647]{Alekzander Kosakowski} \affiliation{Department of Physics
and Astronomy, Texas Tech University, Lubbock, TX 79409, USA}

\author[0000-0002-8655-4308]{A.\ Gianninas} \affiliation{Physics Department, Trinity 
College, 330 Summit Street, Hartford, CT 06106, USA}


\begin{abstract}

	We present the discovery of 17 double white dwarf (WD) binaries from our
on-going search for extremely low mass (ELM) $<0.3$ \msun\ WDs, objects that form
from binary evolution.  Gaia parallax provides a new means of target selection that
we use to evaluate our original ELM Survey selection criteria.  Cross-matching the
Gaia and Sloan Digital Sky Survey (SDSS) catalogs, we identify an additional 36 ELM
WD candidates with $17<g<19$ mag and within the 3-$\sigma$ uncertainties of our
original color selection. The resulting discoveries imply the ELM Survey sample was
90\% complete in the color range $-0.4 < (g-r)_0 < -0.1$ mag (approximately
9,000~K $<$ \teff\ $<$ 22,000~K).  Our observations complete the sample in the SDSS
footprint.  Two newly discovered binaries, J123950.370$-$204142.28 and
J232208.733+210352.81, have orbital periods of 22.5 min and 32 min, respectively,
and are future LISA gravitational wave sources.

\end{abstract}


\section{INTRODUCTION}

	This paper presents new low mass WD observations designed to assess the 
completeness of the ELM Survey.  The ELM Survey is a spectroscopic survey targeting 
extremely low mass $<$0.3~\msun, He-core WDs.  To form such low mass WDs within the 
current age of the Universe requires binary evolution \citep{iben90, marsh95}.  
Different evolutionary pathways can form low mass WDs, and both common-envelope 
and Roche lobe overflow evolution channels contribute to the observed low mass WD 
population \citep{istrate16, sun18, li19}. The ELM Survey is responsible for the 
majority known low mass WDs and the discovery of over 100 detached double degenerate 
binaries \citep{brown20b, brown20, kilic21}.

	Importantly, WD binaries with periods $P\lesssim1$~hr are potential 
multi-messenger sources detectable with both light and gravitational waves with the 
future Laser Interferometer Space Antenna \citep[LISA,][]{amaro17}. Theoretical 
binary population models predict that the majority of multi-messenger WD binaries 
will contain low mass, He-core WDs \citep{korol17, lamberts19, breivik20, thiele21}.  
Observational surveys provide an opportunity to anchor these population models with 
actual numbers.  The difficulty is that observations are based on optical magnitude 
and color, not model parameters like WD mass or binary period.  A complete and 
well-defined observational sample provides the best basis for comparison.

	Gaia astrometry now provides a powerful means to target WDs.  
\citet{fusillo21} use Gaia early Data Release 3 (eDR3) to identify hundreds of 
thousands of high-confidence WDs over the entire sky.  \citet{pelisoli19b} used Gaia 
DR2 to identify candidate low mass WDs.  The original ELM Survey target selection, 
for comparison, was made on the basis of magnitude and color using SDSS passbands.  
Here, we use Gaia astrometry to identify ELM WD candidates in the SDSS footprint 
within 3-$\sigma$ of our original color selection criteria, and present the 
observational results.

	We discuss the details of the target selection in Section 2.  We use 
de-reddened magnitudes and colors indicated with a subscript 0 throughout.  In 
Section 3, we present the observations and derive physical parameters.  These 
measurements complete the color-selected ELM Survey sample in the SDSS 
footprint.  In Section 4, we discuss the ELM Survey completeness and compare it with 
other Gaia-based WD catalogs.  We close by highlighting J1239$-$2041 and J2322+2103, 
which are among the shortest period detached binaries known, and discuss the 
gravitational wave properties of the full sample of 124 WD binaries published in the 
ELM Survey.

\section{ELM WD TARGET SELECTION}

	The original ELM Survey used de-reddened $(u-g)_0$ color as a proxy for 
surface gravity to select candidate low mass WDs.  Normal $\log{g}\sim8$ DA-type WDs 
differ from normal $\log{g}\sim4$ A-type stars by up to 1 magnitude in $(u-g)_0$ 
color in the temperature range $-0.4 < {(g-r)_0} < -0.1$ (about 9,000~K to 
22,000~K) due to the effect of surface gravity on the Balmer decrement in the 
spectrum of the stars \citep{brown12a}.  This region of color space thus provides an 
efficient hunting ground for $\log{g}\sim6$ ELM WDs.

	Color selection does not identify all possible ELM WDs, however, because of 
inevitable confusion with background halo A-type stars and foreground DA WDs with 
similar colors.  During the course of the ELM Survey, we learned that the color 
selection becomes inefficient at $<$9000~K temperatures, or $(g-r)>-0.1$ mag, where 
the hydrogen Balmer lines lose their sensitivity to temperature and gravity.  Low 
mass WDs at $<$9000~K temperatures exist, but subdwarf A-type stars are 100 times 
more common at these temperatures \citep{kepler15, kepler16}.  Subdwarf A-type stars 
are mostly field blue stragglers and metal poor halo stars \citep{brown17a, 
pelisoli17, pelisoli18a, pelisoli18b, pelisoli19a, yu19}.

	Fortunately, ELM WDs also have distinctive $\sim$0.1~\rsun\ radii that fall 
in-between normal $\sim$0.01~\rsun\ DA WDs and $\sim$1~\rsun\ A-type stars.  These 
radii mean ELM WDs are found in a distinct range of absolute magnitude at a fixed 
color (temperature).  Gaia now provides a direct means of selecting ELM WD 
candidates on the basis of parallax.

	To test the ELM Survey completeness for low mass WDs, we select ELM WDs using 
Gaia astrometry in the SDSS footprint as described below.  Our spectroscopic 
observations began with Gaia DR2, and then transitioned to using Gaia eDR3 when it 
became available.  We consider the results of both Gaia data releases.

\subsection{Matching SDSS to Gaia DR2}

	Our initial approach was to cross-match a wide range of stars with possible 
low mass WD colors in SDSS against the Gaia DR2 catalog, so that we properly sample 
the SDSS footprint.  SDSS provides $ugriz$ photometry for about 10,000 deg$^2$ of 
the northern hemisphere sky, mostly located at high Galactic latitudes $|b| \gtrsim 
30\arcdeg$.

	We use SDSS DR16 photometry \citep{ahumada20}, which is identical to SDSS 
DR13 values used by the Gaia Collaboration.  We correct all magnitudes for reddening 
using values provided by SDSS.  Since our WDs have a median distance of 1 kpc, and 
background halo stars are even more distant, applying the full reddening correction 
is important for accurate color selection.

	We began with a simple, broad color cut for blue stars.  We select stars 
from SDSS DR16 with clean photometry (\texttt{clean=1}), $g$-band extinction $<$1 
mag, and colors in the range $-0.5 < (u-g)_0 < 2.0$, $-0.6 < (g-r)_0 < 0.2$, $-0.75 
< (r-i)_0 < 0.25$ mag and errors $<0.15$ mag in $(g-r)_0$ and $<0.2$ mag in 
$(u-g)_0$ and $(r-i)_0$.  The result of our query is 578,883 stars in the magnitude 
range $15<g_0<20$ mag.

	We then performed an exploratory 10 arcsec radius positional cross-match 
with Gaia DR2.  The exploratory cross-match yields 577,253 matches, 16,680 of which 
are duplicates and 10,469 of which have no parallax or proper motion.  Accounting 
for proper motion, we find that a 10.7~yr difference in epoch minimizes the position 
residuals for the blue stars, consistent with the mean epochs of SDSS and Gaia DR2.

	Finally, we perform a clean cross-match between SDSS and Gaia DR2. We 
require \texttt{ruwe}$<$1.4, \texttt{astrometric\_n\_good\_obs\_al}$>$66, and 
\texttt{duplicated\_source}=0 following the guidance of \citet{lindegren18}.  
Our goal is to end up with a clean set of low mass WD candidates, even if the 
\texttt{ruwe} cut may select against wide binaries like HE~0430$-$2457 
\citep{vos18}.  We accept objects that match within 1 arcsec in position, corrected 
for 10.7 years of proper motion, and that match within 1 magnitude between Gaia $G$ 
band and the average SDSS $g+r$ band magnitude.  We reject objects with no parallax 
or proper motion.  This cross-match yields astrometry for 543,534 blue SDSS stars, 
94\% of the original sample.

\subsection{Matching Gaia eDR3 to SDSS}

	Following the release of Gaia eDR3, which has improved completeness for 
faint blue stars and 2$\times$ better astrometric errors than Gaia DR2, we 
transitioned to using Gaia's best neighbour cross-match with SDSS.  This approach 
bases the selection on the Gaia catalog, not on the SDSS catalog, and so imposes an 
unseen astrometric figure of merit to the target selection.  The advantage is that 
the Gaia best neighbour cross-match makes use of the full 5 parameter astrometric 
covariance matrix.

	Starting with the same list of 578,833 blue SDSS stars, we use the 
\texttt{gaiaedr3.sdssdr13\_best\_neighbour} catalog to find Gaia eDR3 matches for 
99.9\% of the stars.  We reject 3,063 objects with no parallax, with matched 
positions that differ by $>$1 arcsec, or with \texttt{number\_of\_mates}$>$0.  We 
then create a clean catalog by applying the \citet{fusillo21} data quality cuts for 
high Galactic latitude white dwarfs\footnote{ 
 \citet{fusillo21} high Galactic latitude data quality cut used:  
 (\texttt{astrometric\_sigma5D\_max} $<$ 1.5 OR (\texttt{ruwe} $\le$ 1.1 AND 
 \texttt{ipd\_gof\_harmonic\_amplitude} $<$ 1)) AND (\texttt{phot\_bp\_n\_obs} $>$ 2 
 AND \texttt{phot\_rp\_n\_obs} $>$ 2) AND 
 ($|$\texttt{phot\_br\_rp\_excess\_factor\_corrected}$|$ $<$ 0.6) AND 
 (\texttt{astrometric\_excess\_noise\_sig} $<$ 2 OR 
 (\texttt{astrometric\_excess\_noise\_sig} $\ge$ 2 AND 
 \texttt{astrometric\_excess\_noise} $<$ 1.5))
	}, excluding the \texttt{parallax\_over\_error} and \texttt{pm\_over\_error} 
criteria.  We will do our own parallax and proper motion selection in the next 
sections.  The final Gaia eDR3 cross-match catalog contains clean astrometry for 
567,769 blue SDSS stars, 98.4\% of the original sample.

	With the Gaia cross-match in hand, we are ready to make various cuts of ELM 
WD candidates beyond the original ELM Survey.  Gaia eDR3 astrometry supersedes 
Gaia DR2 in accuracy and precision, so we use Gaia eDR3 values throughout this 
paper.

\subsection{ELM WD Candidates: Photometric Selection}

	We consider the apparent magnitude range $17<g<19$ mag to maximize our 
volume for finding ELM WDs.  We find that Gaia parallax is reliable to around 19 mag 
for our targets (see below). We adopt $-0.4 < (g-r)_0 < -0.1$ mag to fix the 
temperature range to approximately 9,000~K to 22,000~K and avoid subdwarf A-type 
stars.

\begin{figure} 	
 \includegraphics[width=3.5in]{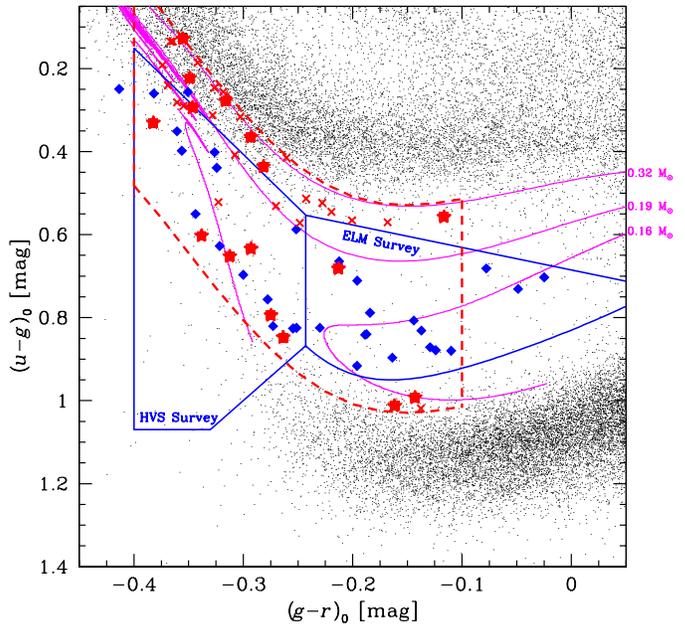}
 \caption{ \label{fig:ugrnew}
	Color-color plot for every star in SDSS with $17<g<19$ mag.  For guidance, 
magenta lines plot theoretical 0.16, 0.19, and 0.32 \msun\ WD tracks 
\citep{althaus13}.  Blue lines mark the approximate color selection of the HVS 
Survey and ELM Survey; blue diamonds mark previously published ELM WD binaries with 
$17<g<19$ mag.  Red dashed lines mark the new color selection; red crosses mark new 
ELM WD candidates, and red stars mark new ELM WD binaries. }
 \end{figure}

	Our first ELM WD selection is based on $(u-g)_0$ color.  For reference, 
Figure \ref{fig:ugrnew} plots $(u-g)_0$ versus $(g-r)_0$ for every star in SDSS with 
$17<g<19$ mag and $E(V-B)<0.1$ mag.  The lower band of dots are normal A-type stars: 
halo blue stragglers and blue horizontal branch stars.  The upper band of dots are 
normal DA-type WDs, stars with the same temperatures but much higher gravities.  
ELM WDs fall in-between.  Magenta lines plot \citet{althaus13} evolutionary tracks 
for 0.16, 0.19, and 0.32 \msun\ WDs (Figure \ref{fig:ugrnew}).  Blue lines mark the 
approximate color-selection of the HVS Survey \citep{brown12b} and the ELM Survey 
\citep{brown12a}, which together formed the basis for our WD observations. Solid 
blue diamonds are the previously published clean ELM sample with $17<g<19$ mag  
minus J0935+4411, a $P=20$~min ELM WD binary that was not selected by color 
\citep{kilic14}.

	The median SDSS $(u-g)_0$ error at $g$=19 mag is 0.038 mag.  Thus to test 
the ELM Survey completeness, we consider the 0.32 \msun\ WD track that is about 0.1 
mag (3-$\sigma$) beyond the original color cut.  A polynomial fit, valid over 
$-0.4<(g-r)_0<-0.1$ mag, yields $(u-g)_0 = 0.535 +1.67(g-r)_0
+24(g-r)_0^2 + 105(g-r)_0^3 +117(g-r)_0^4$.  We use this line as our $(u-g)_0$ color 
cut, bounded by the same line 0.5 mag redder in $(u-g)_0$.  The dashed red line in 
Figure \ref{fig:ugrnew} plots the color selection.  There are 3094 candidates 
$17<g<19$ mag in this color selection region.  The large number of candidates 
reflects the overlap with the main locus of WDs in color space.

\begin{figure} 	
 \includegraphics[width=3.5in]{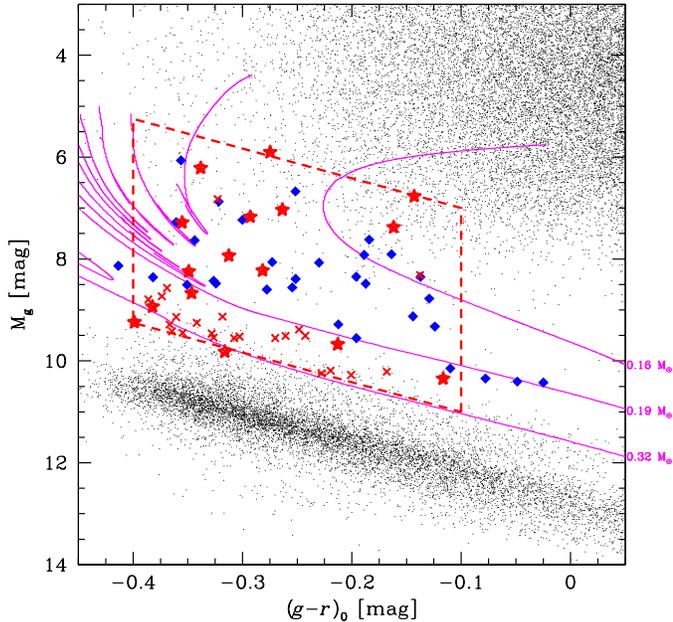}
 \caption{ \label{fig:plxnew}
	Absolute $g$-band magnitude computed from Gaia eDR3 parallax versus 
de-reddened $(g-r)_0$ color for every star in SDSS with $17<g<19$ mag.  Magenta 
lines plot theoretical 0.16, 0.19, and 0.32 \msun\ WD tracks \citep{althaus13}; 
discontinuities are where we clip large excursions due to shell flashes in the 0.19 
\msun\ track for the sake of clarity.  Symbols are the same as before.  Red dashed 
lines mark the Gaia eDR3 selection.  Red stars outside the dashed line were selected 
with Gaia DR2.}
 \end{figure}

\subsection{ELM WD Candidates: Parallax Selection}

	We can also select ELM candidates on the basis of Gaia parallax, independent 
of $(u-g)_0$ color.  Our second approach is to use Gaia parallax to calculate 
absolute $g$-band magnitude $M_g = g_0 - 10 + 5 \log(\pi)$, where $\pi$ is in mas 
and corrected for the global Gaia parallax zero point of $-0.017$ mas 
\citep{lindegren21a}.  We do not bother with higher-order zero point corrections 
because the low mass WD candidates have order-of-magnitude larger $\pm$0.17 mas 
median parallax errors.  We consider the identical magnitude range $17<g<19$ mag, 
temperature range $-0.4 < (g-r)_0 < -0.1$ mag, and require 
\texttt{parallax\_over\_error} $>3$.

	For reference, Figure \ref{fig:plxnew} plots $M_g$ versus $(g-r)_0$ for 
every star in SDSS with $17<g<19$ mag.  The lower band of dots are normal WDs, 
intrinsically faint foreground objects with $M_g \gtrsim 10$ mag.  The upper cloud 
of dots are A spectral-type stars in the halo, intrinsically luminous background 
objects.  ELM WDs fall in-between.  Magenta lines plot the same \citet{althaus13} 
evolutionary tracks for 0.16, 0.19, and 0.32 \msun\ WDs as before (Figure 
\ref{fig:plxnew}).  For the sake of clarity, we clip the shell flash loops in the 
0.19 \msun\ track that briefly jump to high temperature and luminosities.  Blue 
stars mark previously published ELM Survey discoveries with $17<g<19$ mag.

	In Gaia eDR3, the median absolute magnitude error at $g=19$ mag is 0.14 mag 
in our $(u-g)_0$ cut (Figure \ref{fig:ugrnew}).  The median error is substantially 
worse (\texttt{parallax\_over\_error} = 0.3) outside of our $(u-g)_0$ cut because of 
the large number of distant halo stars at similar temperatures.  Thus we consider 
the 0.32 \msun\ track that sits 1 mag below the faintest ELM Survey discovery, the 
same track that we used for the color selection.  A linear fit, valid over 
$-0.4<(g-r)_0<-0.1$ mag, yields $M_g = 11.58 +5.83(g-r)_0$.  We use this line as our 
$M_g$ cut, bounded by the same line 4 mag more luminous (smaller absolute magnitude) 
in $M_g$, as seen by the dashed red line in Figure \ref{fig:plxnew}.  There are 192 
candidates in this $M_g$ selection region, 57 of which are in common with the 
$(u-g)_0$ selection.

\begin{figure} 	
 \includegraphics[width=3.5in]{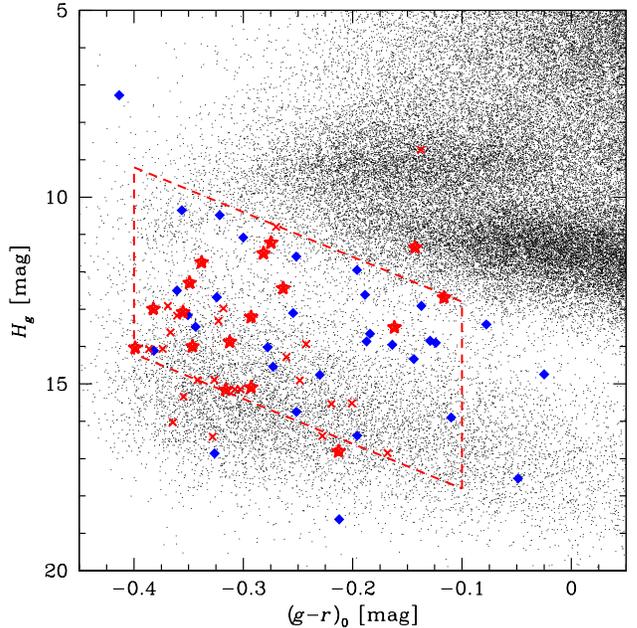}
 \caption{ \label{fig:rpmnew}
	Reduced proper motion $H_g$ computed from Gaia eDR3 proper motions versus 
de-reddened $(g-r)_0$ color for every star in the SDSS with $17<g<19$ mag.  Symbols are 
the same as before.  Red dashed lines mark a possible selection that we do not use 
because of its poor efficiency.}
 \end{figure}

 \begin{splitdeluxetable*}{lcccccccBccccccc}
 \tablecolumns{15}
 \tabletypesize{\scriptsize}
 \tablecaption{Measured and Derived WD Parameters\label{tab:list}}
 \tablehead{
   \colhead{Object} & \colhead{R.A.} & \colhead{Decl.} & \colhead{Pub} & \colhead{Bin} & 
   \colhead{$g_0$} & \colhead{$(g-r)_0$} & \colhead{$(u-g)_0$} & 
   \colhead{$T_{\rm eff}$} & \colhead{$\log g$} & \colhead{mass} & \colhead{$M_g$} & 
   \colhead{$d_{spec}$} & \colhead{plx} & \colhead{Gaia eDR3 Source ID} \\
   \colhead{ } & \colhead{(h:m:s)} & \colhead{(d:m:s)} & \colhead{ } & \colhead{ } & 
   \colhead{(mag)} & \colhead{(mag)} & \colhead{(mag)} & 
   \colhead{(K)} & \colhead{(cm s$^{-2}$)} & \colhead{(\msun)} & \colhead{(mag)} &
   \colhead{(kpc)} & \colhead{(mas)} & \colhead{ } }
	\startdata
J0101$+$0401 &  1:01:28.690 &   4:01:59.00 & 0 & 1 &  $17.219 \pm 0.017$ & $-0.143 \pm 0.033$ & $0.993 \pm 0.035$ &  $ 9284 \pm 120$ & $5.229 \pm 0.089$ & $0.188 \pm 0.013$ & $ 6.74 \pm 0.26$ & $1.245 \pm 0.321$ &  $ 0.7958 \pm 0.0981$ & 2551884196495171712 \\
J0116$+$4249 &  1:16:00.832 &  42:49:38.32 & 0 & 1 &  $18.337 \pm 0.016$ & $-0.293 \pm 0.020$ & $0.634 \pm 0.034$ &  $12968 \pm 180$ & $5.058 \pm 0.045$ & $0.256 \pm 0.028$ & $ 5.07 \pm 0.15$ & $4.506 \pm 0.702$ &  $ 0.5682 \pm 0.1794$ &  373358998783703808 \\
J0130$-$0530 &  1:30:15.918 &  -5:30:25.72 & 0 & 1 &  $18.894 \pm 0.032$ & $-0.293 \pm 0.046$ & $0.365 \pm 0.050$ &  $14727 \pm 120$ & $5.420 \pm 0.025$ & $0.299 \pm 0.053$ & $ 5.47 \pm 0.25$ & $4.834 \pm 1.233$ &  $-0.0249 \pm 0.2712$ & 2479121368826845568 \\
J0745$+$2104 &  7:45:00.527 &  21:04:31.37 & 0 & 1 &  $18.578 \pm 0.013$ & $-0.399 \pm 0.025$ & $0.016 \pm 0.027$ &  $22614 \pm 290$ & $7.329 \pm 0.041$ & $0.397 \pm 0.016$ & $ 9.21 \pm 0.08$ & $0.747 \pm 0.059$ &  $ 1.3408 \pm 0.2521$ &  672983820089822464 \\
J0806$-$0716 &  8:06:50.022 &  -7:16:36.11 & 0 & 1 &  $18.065 \pm 0.011$ & $-0.312 \pm 0.016$ & $0.652 \pm 0.031$ &  $12374 \pm 160$ & $6.109 \pm 0.041$ & $0.203 \pm 0.027$ & $ 8.01 \pm 0.16$ & $1.027 \pm 0.167$ &  $ 0.9263 \pm 0.1802$ & 3063885736023355904 \\
J0820$+$4543 &  8:20:10.339 &  45:43:01.70 & 0 & 1 &  $17.924 \pm 0.023$ & $-0.316 \pm 0.030$ & $0.278 \pm 0.033$ &  $17356 \pm 220$ & $7.458 \pm 0.039$ & $0.412 \pm 0.016$ & $ 9.98 \pm 0.07$ & $0.388 \pm 0.029$ &  $ 2.3783 \pm 0.1751$ &  928585329693880832 \\
J1239$-$2041 & 12:39:50.370 & -20:41:42.28 & 0 & 1 &  $18.621 \pm 0.021$ & $-0.346 \pm 0.027$ & $0.294 \pm 0.047$ &  $17575 \pm 260$ & $6.939 \pm 0.046$ & $0.291 \pm 0.013$ & $ 9.04 \pm 0.13$ & $0.824 \pm 0.107$ &  $ 1.0068 \pm 0.2309$ & 3503613283880705664 \\
J1313$+$5828 & 13:13:49.976 &  58:28:01.39 & 0 & 1 &  $18.380 \pm 0.019$ & $-0.382 \pm 0.025$ & $0.331 \pm 0.028$ &  $16610 \pm 230$ & $6.938 \pm 0.043$ & $0.271 \pm 0.015$ & $ 9.22 \pm 0.11$ & $0.678 \pm 0.073$ &  $ 1.2747 \pm 0.1269$ & 1566990986557895040 \\
J1632$+$4936 & 16:32:42.394 &  49:36:14.60 & 0 & 1 &  $17.946 \pm 0.017$ & $-0.162 \pm 0.023$ & $1.013 \pm 0.031$ &  $ 9156 \pm 120$ & $5.746 \pm 0.064$ & $0.269 \pm 0.021$ & $ 7.71 \pm 0.16$ & $1.117 \pm 0.176$ &  $ 0.7530 \pm 0.0896$ & 1411455519097674368 \\
J1906$+$6239 & 19:06:00.874 &  62:39:23.71 & 0 & 1 &  $17.622 \pm 0.013$ & $-0.338 \pm 0.018$ & $0.603 \pm 0.028$ &  $13570 \pm 180$ & $5.341 \pm 0.042$ & $0.259 \pm 0.040$ & $ 5.67 \pm 0.20$ & $2.460 \pm 0.495$ &  $ 0.5060 \pm 0.0988$ & 2252265701675503616 \\
J2104$+$1712 & 21:04:03.842 &  17:12:32.17 & 0 & 1 &  $18.160 \pm 0.025$ & $-0.213 \pm 0.027$ & $0.682 \pm 0.038$ &  $ 8927 \pm 110$ & $6.561 \pm 0.048$ & $0.183 \pm 0.010$ & $10.40 \pm 0.11$ & $0.357 \pm 0.041$ &  $ 1.9920 \pm 0.1495$ & 1764073394954901504 \\
J2149$+$1506 & 21:49:11.107 &  15:06:37.71 & 0 & 1 &  $18.065 \pm 0.021$ & $-0.355 \pm 0.029$ & $0.128 \pm 0.036$ &  $21164 \pm 300$ & $6.595 \pm 0.044$ & $0.267 \pm 0.032$ & $ 7.95 \pm 0.17$ & $1.055 \pm 0.181$ &  $ 0.6810 \pm 0.1842$ & 1769307791858602880 \\
J2151$+$2730 & 21:51:11.472 &  27:30:14.45 & 0 & 1 &  $17.016 \pm 0.012$ & $-0.275 \pm 0.016$ & $0.795 \pm 0.025$ &  $11901 \pm  90$ & $5.257 \pm 0.028$ & $0.189 \pm 0.010$ & $ 6.07 \pm 0.11$ & $1.546 \pm 0.170$ &  $ 0.5820 \pm 0.0763$ & 1800183147814621056 \\
J2257$+$3023 & 22:57:02.141 &  30:23:38.50 & 0 & 1 &  $18.343 \pm 0.015$ & $-0.117 \pm 0.020$ & $0.558 \pm 0.030$ &  $ 9947 \pm 120$ & $7.324 \pm 0.043$ & $0.334 \pm 0.016$ & $11.13 \pm 0.08$ & $0.277 \pm 0.023$ &  $ 2.5042 \pm 0.1695$ & 1886328505164612864 \\
J2306$+$0224 & 23:06:37.879 &   2:24:29.61 & 0 & 1 &  $16.910 \pm 0.021$ & $-0.264 \pm 0.025$ & $0.848 \pm 0.055$ &  $11211 \pm 180$ & $5.473 \pm 0.050$ & $0.201 \pm 0.015$ & $ 6.69 \pm 0.14$ & $1.105 \pm 0.159$ &  $ 1.0422 \pm 0.0987$ & 2658636742508253568 \\
J2322$+$2103 & 23:22:08.733 &  21:03:52.81 & 0 & 1 &  $18.616 \pm 0.019$ & $-0.349 \pm 0.026$ & $0.223 \pm 0.037$ &  $16677 \pm 220$ & $6.765 \pm 0.041$ & $0.250 \pm 0.021$ & $ 8.88 \pm 0.16$ & $0.884 \pm 0.141$ &  $ 0.8261 \pm 0.2503$ & 2825856243296564608 \\
J2348$+$2804 & 23:48:52.300 &  28:04:38.41 & 0 & 1 &  $18.636 \pm 0.019$ & $-0.281 \pm 0.028$ & $0.436 \pm 0.036$ &  $13906 \pm 180$ & $6.204 \pm 0.041$ & $0.220 \pm 0.037$ & $ 7.96 \pm 0.23$ & $1.365 \pm 0.308$ &  $ 0.8113 \pm 0.2267$ & 2866648537004311808 \\
	\enddata
	\tablecomments{
	Classifications are set to 1 if true or 0 if false, i.e., Pub=1 indicates a 
previously published object, and Bin=1 indicates binaries with orbital solutions.
	J0130$-$0530, J0820+4543, and J2151+2730 were selected with Gaia DR2 and do 
not belong in the Gaia eDR3-based sample, but are ELM WD binaries included here for 
completeness.
	We list parameters for the 17 new binaries to illustrate the content; the 
table is available in its entirety in machine-readable form.}
 \end{splitdeluxetable*}

\subsection{ELM Candidates: optional Reduced Proper Motion Selection}

	Parallax provides a powerful but not a perfect selection criteria.  Given 
Gaia measurement errors, WDs more distant than about 1 kpc become confused with halo 
stars on the basis of parallax alone.  Proper motion provides an independent 
constraint.  Given that halo stars are an order of magnitude more distant than ELM 
WDs, we can use reduced proper motion as an optional third selection criteria.

	We compute $g$-band reduced proper motion $H_g = g_0 + 5\log{\mu} - 10$, 
where $\mu$ is the Gaia proper motion in \mas.  In Gaia eDR3, the median reduced 
proper motion error at $g=19$ mag is 6\% within our color cut.  The precision is 
deceptive, however, as the intrinsic velocity dispersion of stars causes overlap in 
reduced proper motion space.  

	For reference, Figure \ref{fig:rpmnew} plots $H_g$ versus $(g-r)_0$ for 
every star in SDSS with $17<g<19$ mag.  The lower cloud of dots are normal WDs, 
nearby objects that exhibit larger proper motions than more distant
stars of the same temperature.  The upper clouds of dots are A spectral-type stars 
in the halo.  ELM WDs fall approximately in-between the foreground WDs 
and background halo stars, but with considerable spread.
	The problem is that known ELM WDs are a mix of disk and halo populations, 
and so exhibit a large spread in reduced proper motion space.

	We initially considered an empirical reduced proper motion selection $14 + 
12(g-r)_0 < H_g < 19 + 12(g-r)_0$ that includes previously published discoveries 
and minimizes the contribution of background halo stars (Figure \ref{fig:rpmnew}).  
There are 5267 candidates in this $H_g$ selection.  
However, the $H_g$ selection has substantial overlap with the population of normal 
WDs, and legitimate ELM WDs are excluded.  Given the inefficiency of the $H_g$ 
selection compared to the other selection criteria, we do not consider it further.

\subsection{Final Selection}

	Parallax selection provides the most efficient means for identifying low 
mass WDs candidates with minimal contamination from foreground and background 
sources, followed by color selection.  We combine the parallax and color selection 
to optimize our observing program.

	Fifty seven objects jointly satisfy the Gaia eDR3 parallax and SDSS color 
selection criteria with $17<g<19$ mag.  These objects are the highest-probability 
ELM WD candidates in the SDSS footprint.  Table~\ref{tab:list} presents the list.  
We include 3 additional objects (J0130$-$0530, J0820+4543, and J2151+2730) in 
Table~\ref{tab:list} that were selected and observed using Gaia DR2 but do not 
satisfy the Gaia eDR3 parallax cut (i.e., the 3 objects that fall just outside 
selection region in Figure \ref{fig:plxnew}). We include them here for completeness.

	Twenty one objects are previously published \citep{brown20b, brown20} and 
labeled Pub=1 in Table~\ref{tab:list}.  We present spectroscopy for the 
remaining targets in this paper, plus orbital solutions for 17 new binary systems.  
One of the new binary systems, J1239$-$2041, was identified as an ELM white dwarf
by \citet{kosakowski20}, but its orbital parameters were unconstrained.
Binaries are labeled Bin=1 in Table \ref{tab:list}.

\section{OBSERVATIONS}

\subsection{Spectroscopy}

	Our observational strategy is to obtain a spectrum for each target, and then 
follow-up likely ELM WDs with time-series spectroscopy.  We began observations in 
late 2018 after the release of Gaia DR2.  Follow-up spectroscopy started a year 
later, in late 2019.  We lost the 2020 observing seasons to telescope closures.  
Observations resumed the second half of 2021.

	We obtained spectra for all 39 of the previously unobserved objects at the 
6.5m MMT telescope with the Blue Channel spectrograph \citep{schmidt89}.  We 
configured the spectrograph with the 832 l~mm$^{-1}$ grating and the 1.25\arcsec\ 
slit to provide 3550 -- 4500 \AA\ coverage at 1.2 \AA\ spectral resolution.  
Spectra are flux-calibrated using a nightly standard star observation.  All 
observations are paired with a comparison lamp spectrum for accurate wavelength 
calibration.

	For 7 southern targets, we obtained additional spectra at the 6.5m Magellan 
telescope with the MagE spectrograph and at the 4.1m SOAR telescope with the 
Goodman High Throughput spectrograph \citep{clemens04} to improve the time-baseline 
for radial velocity variables J0806$-$0716 and J1239$-$2041 and to validate the 
stellar atmosphere fits for the other five southern targets. At Magellan we used the
0.85 arcsec slit, providing a resolving power of R = 4800. For SOAR observations, we
used the blue camera with the 1 arcsec slit and the 930 lines mm$^{-1}$ grating
resulting in a spectral resolution of 2.5 \AA. SOAR observations were obtained
as part of the NOAO Programs 2019A-0134 and 2021A-0007. 

	We also obtained time-series spectroscopy of J0101+0401 at the Apache Point 
Observatory 3.5m telescope with the Dual Imaging Spectrograph (DIS). We used the 
B1200 and R1200 gratings with a 1.5 arcsec slit for a dispersion of 0.6 \AA\ per 
pixel.
 
	We measure radial velocities by cross-correlating the observations using the 
RVSAO package \citep{kurtz98} with high signal-to-noise templates of similar 
stellar parameters and known velocities obtained in the identical spectrograph 
set-up.  Any systematic observational errors (i.e., in the wavelength solution) 
are common to the targets and the templates, allowing us to measure precise relative 
velocities.  The radial velocities typically have 10 to 15 \kms\ precision.

\begin{figure} 	
 \includegraphics[width=3.5in]{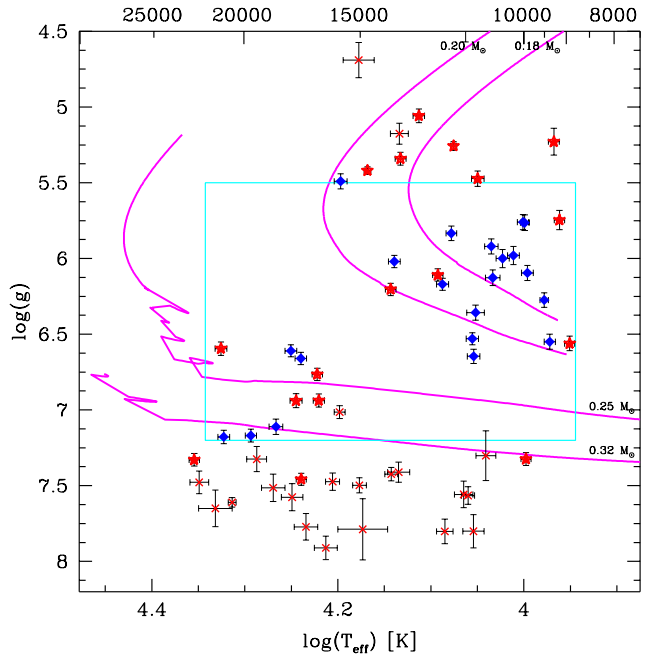}
 \caption{ \label{fig:teff}
	Effective temperature versus surface gravity for the final sample of 57 + 3 
low mass WD candidates.  Cyan box marks our clean ELM WD sample, the region that 
maximizes the overlap between the data and the ELM WD tracks \citep[magenta 
lines,][]{istrate16}.  Symbols same as before.  }
 \end{figure}

\subsection{Stellar Atmosphere Fits}

	We measure the effective temperature and surface gravity of each target by 
fitting pure hydrogen-atmosphere models to its summed, rest-frame spectrum.  The 
model grid spans 4000 $<$ \teff\ $<$ 35,000~K and 4.5 $<$ \logg\ $<$ 9.5 
\citep{gianninas11, gianninas14b, gianninas15} and includes the Stark broadening 
profiles from \citet{tremblay09}.  We apply the \citet{tremblay15} 3D stellar 
atmosphere model corrections, which are relevant to $\lesssim10,000$~K objects, when 
needed.  We present the corrected stellar atmosphere parameters for all targets in 
Table \ref{tab:list}.

	Figure \ref{fig:teff} plots \teff\ and \logg\ for the sample of 57 + 3 
objects, and \citet{istrate16} evolutionary tracks for 0.18, 0.20, 0.25, and 0.32 
\msun\ WDs for comparison.  We adopt \citet{istrate16} models for the final ELM 
WD analysis because the tracks are computed for both disk and halo progenitors.  We 
see both disk and halo objects in the observations, as discussed below.  The cyan 
box is the region we refer to as the clean ELM WD sample: the region that maximizes 
the overlap between the ELM WD models and the data.

	Figure \ref{fig:teff} shows that our choice of $-0.4 < (g-r)_0 < -0.1$ mag 
color is effective in selecting 9000 $<$ \teff\ $<$ 22,000~K objects and excluding 
stars at cooler temperatures.  New objects (red symbols) are mostly found at both 
higher and lower \logg\ compared to the original ELM Survey, consistent with the 
expanded color selection, but nine of the new objects are found in the clean ELM 
region.  We explore these objects further below.

	Interestingly, the overall distribution of points in Figure \ref{fig:teff} 
does not uniformly fill \teff\ and \logg\ space. The onset of hydrogen shell flashes 
in the WD evolutionary tracks around $\sim$0.2 \msun\ creates a gap in the tracks in 
\teff\ and \logg\ space.  Observed WDs largely overlap the WD evolutionary tracks in 
\teff\ and \logg\ space, and do not uniformly fill the plot.

	The joint parallax and color criteria are very efficient at finding ELM WDs:  
half (29) of our targets fall in the clean ELM WD region that directly overlap the 
ELM WD tracks in the range $5.5<\log{g}<7.2$.

\subsection{Disk and Halo Populations}

	Before selecting an evolutionary track for the analysis, we must first 
assess whether an observed target is a disk or halo object.  We use 3 dimensional 
space velocities to make this assessment.  In brief, we compute space velocities 
using our systemic radial velocities, corrected for the gravitational redshift of 
the observed WD, and Gaia proper motions. We then compare the velocity of the low 
mass WD with the velocity ellipsoid of the disk and halo as described in 
\citet{brown20}.  The result is essentially the same as found by \citet{brown20}:  
70\% of the observed objects have motions consistent with the thick disk velocity 
ellipsoid and 30\% have motions consistent with the halo.

\begin{figure} 	
 \includegraphics[width=3.5in]{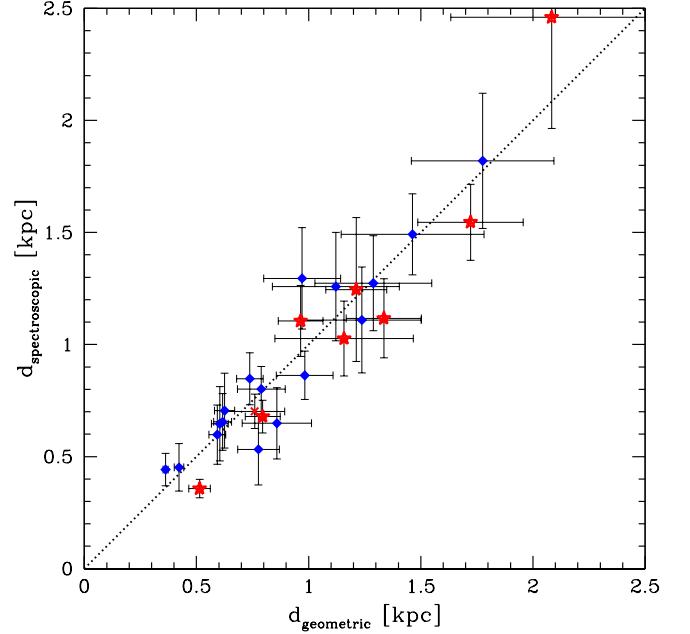}
 \caption{ \label{fig:dtest}
	Spectroscopic distance estimates derived from \citet{istrate16} ELM WD 
tracks compared to geometric distances measured from Gaia eDR3.  Symbols same as 
before.  Dotted line is the 1:1 ratio.}
 \end{figure}

 \begin{deluxetable*}{lccccc}	
 \tablecolumns{6}
 \tabletypesize{\scriptsize}
 \tablecaption{Binary Parameters\label{tab:binaries}}
 \tablehead{
   \colhead{Object} & \colhead{$P$} & \colhead{$k$} & \colhead{$\gamma$} &
	\colhead{$M_{\rm 2,min}$} & \colhead{$\log\tau_{\rm GW,max}$} \\
   \colhead{ } & \colhead{(days)} & \colhead{(\kms)} & \colhead{(\kms)} &
	\colhead{(\msun)} & \colhead{(yr)} }
	\startdata
J0101$+$0401 & $0.18332 \pm 0.00284$ & $199.5 \pm  7.1$ & $ -11.3 \pm  6.9$ & $>$0.35 & $<$9.8  \\
J0116$+$4249 & $0.33400 \pm 0.00015$ & $237.8 \pm  4.6$ & $-204.7 \pm  3.1$ & $>$0.81 & $<$10.1  \\
J0130$-$0530 & $0.63648 \pm 0.00072$ & $191.2 \pm  5.7$ & $ -29.3 \pm  3.9$ & $>$0.85 & $<$10.8  \\
J0745$+$2104 & $0.53964 \pm 0.00511$ & $132.2 \pm  4.6$ & $  55.0 \pm  3.6$ & $>$0.46 & $<$10.7  \\
J0806$-$0716 & $0.70555 \pm 0.01554$ & $170.7 \pm 11.2$ & $  26.9 \pm 13.6$ & $>$0.63 & $<$11.1  \\
J0820$+$4543 & $0.31553 \pm 0.00042$ & $153.1 \pm  3.7$ & $  86.5 \pm  3.1$ & $>$0.44 & $<$10.1  \\
J1239$-$2041 & $0.01563 \pm 0.00013$ & $557.2 \pm 10.4$ & $  -8.9 \pm  7.5$ & $>$0.61 & $<$6.6  \\
J1313$+$5828 & $0.07395 \pm 0.00018$ & $321.7 \pm  6.5$ & $ -21.3 \pm  4.8$ & $>$0.56 & $<$8.5  \\
J1632$+$4936 & $0.10141 \pm 0.00016$ & $209.7 \pm  7.2$ & $ -16.1 \pm  6.7$ & $>$0.33 & $<$9.0  \\
J1906$+$6239 & $0.32939 \pm 0.00005$ & $271.2 \pm  3.0$ & $ -57.8 \pm  3.2$ & $>$1.06 & $<$10.0  \\
J2104$+$1712 & $0.23750 \pm 0.00022$ & $286.6 \pm  6.0$ & $   7.6 \pm  5.2$ & $>$0.86 & $<$9.8  \\
J2149$+$1506 & $0.08541 \pm 0.00016$ & $290.3 \pm 12.0$ & $ -31.2 \pm  4.7$ & $>$0.51 & $<$8.7  \\
J2151$+$2730 & $0.51593 \pm 0.00316$ & $203.9 \pm  6.7$ & $  31.0 \pm 10.3$ & $>$0.72 & $<$10.8  \\
J2257$+$3023 & $0.13489 \pm 0.00016$ & $226.3 \pm  3.2$ & $   2.1 \pm  2.8$ & $>$0.47 & $<$9.1  \\
J2306$+$0224 & $0.28728 \pm 0.00009$ & $148.3 \pm  5.7$ & $ -11.5 \pm  4.2$ & $>$0.28 & $<$10.4  \\
J2322$+$2103 & $0.02220 \pm 0.00025$ & $248.1 \pm  4.3$ & $   7.8 \pm  3.3$ & $>$0.19 & $<$7.5  \\
J2348$+$2804 & $0.92013 \pm 0.01532$ & $ 89.3 \pm 12.2$ & $  30.4 \pm 11.8$ & $>$0.25 & $<$11.7  \\
	\enddata
	\tablecomments{J0745+2104 observations have a period alias at $P=0.343$ days.}
  \end{deluxetable*}

\subsection{White Dwarf Parameters}

	We then estimate WD mass and luminosity by interpolating the measured \teff\ 
and \logg\ through WD evolutionary tracks.  For 34 objects with $\log{g}<7.2$, we 
use \citet{istrate16} tracks with rotation and diffusion.  We apply $Z=0.02$ tracks 
to objects with disk kinematics, and $Z=0.001$ tracks to objects with halo 
kinematics. We then estimate distances from the de-reddened SDSS apparent magnitude 
and the absolute magnitude derived from the tracks, $d_{spec} = 
10^{((g_0-M_g)/5-2)}$~kpc.

	Figure \ref{fig:dtest} compares our $d_{spec}$ estimates derived from 
\citet{istrate16} tracks against the $d_{geo}=1/\pi$ distances measured 
geometrically by Gaia.  We plot the 29 objects with distance ratio uncertainties 
better than 30\%.  The mean ratio is $d_{spec}/d_{geo}=1.02\pm0.04$.  Thus the 
\citet{istrate16} tracks provide a remarkably accurate measure of mass and radius 
for ELM WDs.

	The most discrepant point in Figure \ref{fig:dtest} is one of the new ELM WD
binaries, J2104+1712, with $d_{spec}/d_{geo}=0.69\pm0.15$.  J2104+1712 is a $P=5.7$
hr binary.  Our spectra show no evidence for a double-lined spectroscopic system or
emission in the cores of the Balmer lines, however it is possible that such effects
could artificially inflate our \logg\ measurement.  If the actual \logg\ were 0.5
dex lower (the WD had larger radius), the derived absolute magnitude would be
$\sim$1 mag more luminous for the same temperature and bring the spectroscopic
distance into agreement with the geometric distance.  

	For the remaining objects that do not overlap the \citet{istrate16} tracks, 
we estimate WD parameters using \citet{althaus13} helium-core WD tracks for objects 
up to $\log{g}=7.5$, and \citet{tremblay11} WD tracks for objects above 
$\log{g}=7.5$.  We note that $d_{spec}$ estimates for the high-gravity WDs are 
systematically smaller than the Gaia distances, very similar to J2104+1712, but we 
defer a detailed investigation.  We focus our attention on the well-measured ELM 
WDs.

\begin{figure} 	
 \includegraphics[width=3.5in]{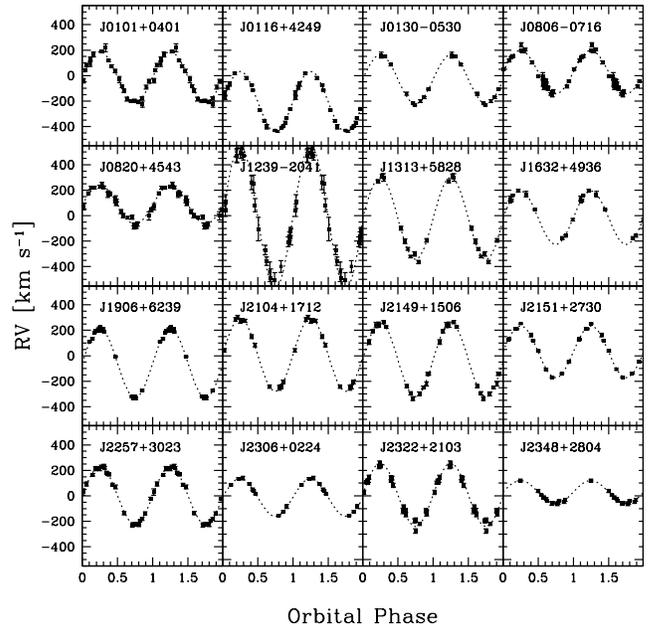}
 \caption{ \label{fig:orb}
	Radial velocities phased to the best fit orbital period for the new 
ELM WD binaries.  The measurements for all 17 binaries are available in the 
electronic Journal as the Data behind the Figure. }
 \end{figure}

\subsection{Binary Orbital Solutions}

	Follow-up time-series spectroscopy reveals significant 200 \kms\ to 1100 
\kms\ radial velocity variability in every $5<\log{g}<7.2$ target.  Velocity 
variability in $\log{g}>7.5$ targets is not significant given our observational 
errors.  For the new targets with $5<\log{g}<7.2$, we acquire a total of 10 to 20 
spectra spaced over a few different nights.  This sampling is sufficient to 
determine binary orbital elements for 14 targets, plus another 3 targets at slightly 
higher \logg.  Only one object in the clean ELM region, J1240$-$0958, lacks 
sufficient coverage to determine its orbital parameters.

	Table \ref{tab:binaries} presents the 17 binaries, and Figure \ref{fig:orb} 
presents the radial velocities phased to the best fit orbital solution.  Our 
computational approach is to minimize $\chi^2$ for a circular orbit fit to each set 
of radial velocities.  We estimate errors by re-sampling the radial velocities with 
their errors and re-fitting the orbital solution 10,000 times.  We report the 15.9\% 
and 84.1\% percentiles of the parameter distributions in Table \ref{tab:binaries}.  
Systemic velocities are corrected for gravitational redshift using the WD parameters 
in Table \ref{tab:binaries}.  One target, J0745+2104, has a significant period 
alias, and therefore is excluded from Figure \ref{fig:orb}.

\subsection{Time Series Photometry of J1239$-$2041}

	J1239$-$2041 is the shortest period ($P=22.5$ min) binary in the new 
discoveries, and among the shortest period detached WD binaries known.  
Ultra-compact $P<1$ hr WD binaries commonly exhibit ellipsoidal variation, 
reflection effects, and eclipses in their light curves, all of which constrain the 
inclination of the binary \citep[e.g.][]{hermes14,burdge20a}.

\begin{figure} 	
 \includegraphics[width=3.5in]{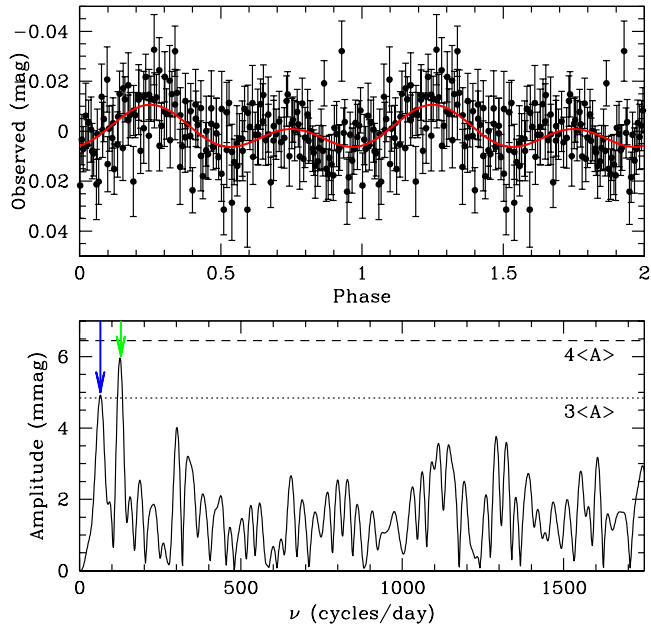}
 \caption{ \label{fig:lc}
	Gemini $g$-band light curve of J1239$-$2041 (top panel) folded at the 
orbital period, and repeated for clarity. The solid red line displays our 
best-fitting model. The bottom panel shows the Fourier transform of this light 
curve, and the 3$\langle {\rm A}\rangle$ and 4$\langle {\rm A}\rangle$ levels, where 
$\langle {\rm A}\rangle$ is the average amplitude in the Fourier Transform. 
Blue and green arrows mark the 1$\times$ and 2$\times$ orbital frequency. }
 \end{figure}

	We used Gemini South GMOS to acquire time-series imaging of J1239$-$2041 on 
UT 2021 June 28 as part of the program GS-2021A-DD-108.  We obtained 161 
back-to-back $g$-band exposures over a 66 min baseline to test for 
photometric variability.  The short time-series is sufficient to rule out the 
presence of eclipses (Figure \ref{fig:lc}).

	A frequency analysis finds marginally significant photometric variability at 
the radial velocity orbital period.  We perform a simultaneous, non-linear 
least-squares fit that includes the amplitude of the Doppler beaming ($\sin{\phi}$), 
ellipsoidal variations ($\cos{2\phi}$), and reflection ($\cos{\phi}$).  The best-fit 
model, plotted as the red line in Figure \ref{fig:lc}, has relativistic beaming = 
$0.45 \pm 0.12$\%, ellipsoidal variation = $0.53 \pm 0.11$\%, and reflection effect 
= $0.00^{+0.04}_{-0.0}$\%.

\begin{figure*} 	
 \includegraphics[width=6.5in]{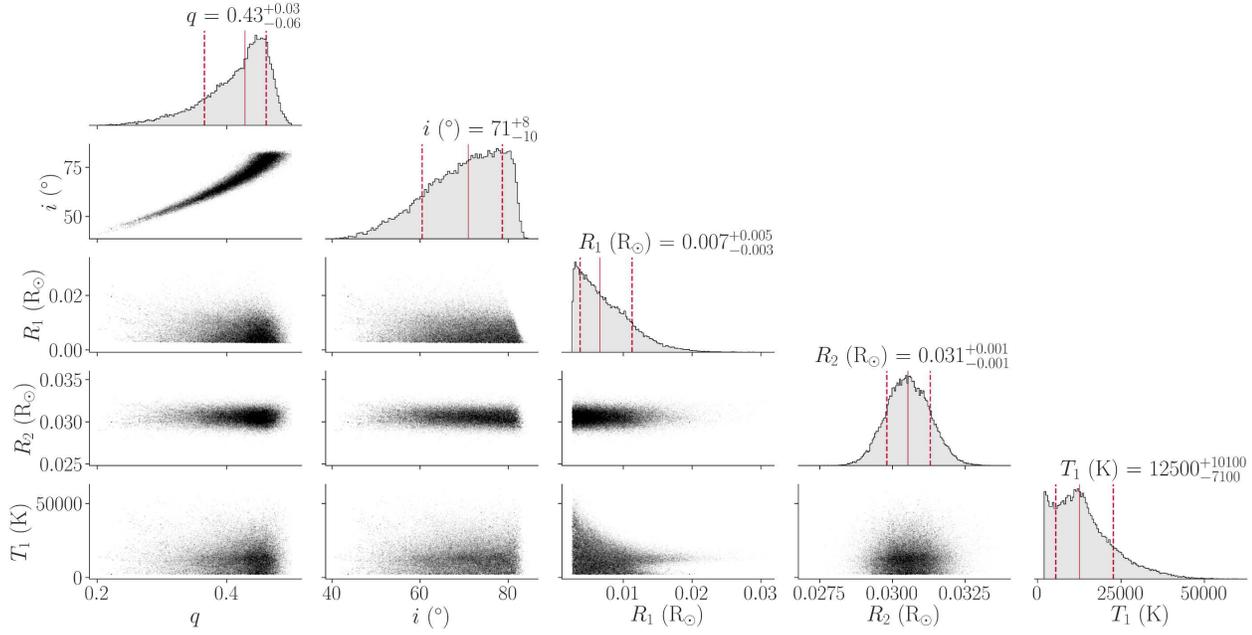}
 \caption{ \label{fig:lcurve}
	Corner plot showing the parameter distributions from LCURVE.  The unseen 
companion parameters are labeled $T_1$ and $R_1$.  The ELM WD radius, labeled $R_2$, 
matches the input priors. The inclination constraint is $i=71^{+8}_{-10}$\arcdeg.}
 \end{figure*}

	We also use LCURVE to fit the mass ratio, scaled radii, effective 
temperatures, and velocity scale used to account for Doppler beaming and 
gravitational lensing.  We use fixed quadratic limb-darkening values and fixed 
gravity-darkening values \citep{claret20} based on our spectroscopic 
atmospheric parameters for the ELM WD and a putative 9000~K, \logg=8 companion.  
We place Gaussian priors on the ELM WD temperature, mass, and velocity 
semi-amplitude based on our spectroscopic and radial velocity measurements.  Our 
MCMC algorithm uses 256 walkers, each performing 1000 fit iterations.  We remove the 
first 100 of each as burn-in, leaving 230,400 iterations from which we created 
the final parameter distributions plotted in Figure \ref{fig:lcurve}.  Note 
that LCURVE refers to the parameters of the unseen companion -- the more massive 
star -- with a subscript 1 and the ELM WD with a subscript 2.

	We derive a $0.68^{+0.11}_{-0.06}$~\msun\ companion mass given the fitted 
mass ratio and our spectroscopic $0.291\pm0.013$ \msun\ ELM WD mass.  As seen in 
Figure \ref{fig:lcurve}, the companion's $12500^{+10100}_{-7100}$~K temperature and 
$0.007^{+0.005}_{-0.003}$~\rsun\ radius are poorly constrained by our short 
time-series observations, but the parameters are consistent with WD tracks.  LCURVE 
returns a radius of $0.031^{+0.001}_{-0.001}$~\rsun\ for the ELM WD in perfect 
agreement with input priors.

	The lack of eclipses limits the inclination of J1239$-$2041 to 
$i<81\arcdeg$.  However, the 0.53\% ellipsoidal amplitude implies an approximately 
edge-on orientation. The LCURVE fit yields $i=71^{+8}_{-10}$\arcdeg.  We also 
detect Doppler beaming, and the predicted 0.38\% beaming amplitude is consistent 
within the errors of the observed 0.45\% value.  A longer baseline light curve is 
needed to more accurately constrain J1239$-$2041's parameters.

\section{DISCUSSION}

\subsection{Observational Completeness for ELM WDs}

	A primary goal of this paper is to assess the observational completeness of 
the published ELM Survey sample.  We search for probable low mass WD candidates by 
expanding the original color selection region by the 3-$\sigma$ SDSS color 
uncertainties, and then applying Gaia parallax constraints to identify 57 candidates 
with $17<g<19$ mag across the entire SDSS footprint in the color range 
$-0.4<(g-r)_0<-0.1$ mag.  

	Follow-up observations are complete.  Twenty one candidates were previously 
published, and observations for the remaining 36 candidates are presented in the 
previous section.  All but one of the 57 candidates are WDs. (We 
consider J2338+1235 a non-WD on the basis of its spectroscopic \logg\ = 
$4.44\pm0.13$.  J2338+1235 has one of the poorest parallax constraints in our 
sample, \texttt{parallax\_over\_error} = 3.8, and its $-299\pm11$ \kms\ heliocentric 
radial velocity suggests it is a halo star.)

	In terms of target selection, the ELM Survey missed a dozen low mass WD 
candidates with $-0.4<(g-r)_0<-0.1$ mag in SDSS (Figure \ref{fig:ugrnew}).  
Nearly all of the missing candidates are in blue half defined by the HVS survey 
\citep{brown12a} and are missing because they failed the selection criteria at the 
time of observation.  The HVS Survey was executed prior to SDSS DR8 when the SDSS 
photometric calibration and sky coverage changed every data release.  Candidates 
near the hot edges of the color selection in Figure \ref{fig:ugrnew} were 
particularly sensitive to the changing calibrations.  J1313+5828 had $(g-r)_0<-0.40$ 
in DR7 and earlier data releases, for example, and so was not included in the 
original sample.  Five candidates located at southern Galactic latitudes or low 
declination did not appear in SDSS sky coverage until DR8.  Four more candidates 
were actually observed by the HVS Survey, but were not included in ELM Survey 
follow-up because the first spectra indicated $\log{g}\le5$ surface gravity and the 
objects were not considered probable WDs.

	In terms of sample completeness for ELM WDs, we estimate that the published 
ELM Survey was 90\% complete.  Our observations identify a total of 9 new ELM WDs, 
10 if you count J2306+0224 just outside the clean ELM WD region in Figure 
\ref{fig:teff}. Only three of the new ELM WDs fall in the original color selection 
criteria:  a 9\% increase in the clean sample of ELM WDs with $17<g<19$ mag and 
$-0.4<(g-r)_0<-0.1$ mag.  This color-selected sample should now be complete over 
the entire SDSS footprint.

	Of course ELM WDs exist outside our original color selection criteria, as 
demonstrated by the broader selection used in this paper.  Yet most of the 
candidates from the expanded color selection in Figure \ref{fig:ugrnew} are normal 
WDs.  \citet{torres22} estimate that only 1\% of all WDs are He-core objects based 
on the Gaia 100 pc sample.  Applying a joint parallax constraint is thus important 
for identifying real ELM WDs in a crowded region of parameter space. It is 
interesting to explore how other Gaia-based catalogs of WDs and ELM WD candidates 
compare.

\begin{figure} 	
 \includegraphics[width=3.5in]{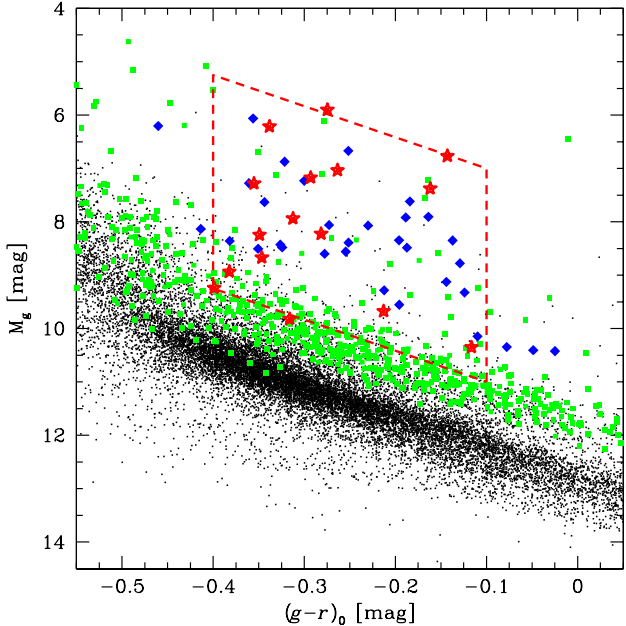}
 \caption{ \label{fig:plxcomp}
	Comparison of our clean ELM WD sample (blue and red symbols, same as Figure 
\ref{fig:plxnew}), high-confidence WDs from the \citet{fusillo21} Gaia eDR3 WD 
catalog (black dots), and objects from the \citet{pelisoli19b} Gaia DR2 ELM WD 
candidate catalog (green squares).  To make a fair comparison, we plot only those 
objects within SDSS with $17<g<19$ mag.  }
 \end{figure}

\subsubsection{Comparison with Gaia WD catalogs}

	Figure \ref{fig:plxcomp} compares the clean ELM WD sample with the 
\citet{fusillo21} Gaia eDR3 WD catalog and the \citet{pelisoli19b} Gaia DR2 ELM WD 
candidate catalog.  We restrict the figure to objects within SDSS with $17<g<19$ mag 
to make a one-to-one comparison.

	\citet{fusillo21} select 1.3 million WD candidates from the Gaia eDR3 
catalog from which they identify 390,000 high-confidence $P_{\rm WD}>0.75$ sources.  
The 21,500 high-confidence sources with SDSS $17<g<19$ mag are drawn as black 
dots in Figure \ref{fig:plxcomp}.  The majority of sources fill the locus of normal 
WDs. 

	Interestingly, the \citet{fusillo21} WD catalog includes 58 of our 60 
objects, and two-thirds (38) are high-confidence WDs.  The \citet{fusillo21} 
photometric hydrogen atmosphere parameter estimates are not very reliable for our 
class of objects, however, on average $7\pm16\%$ lower in \teff\ and $0.4\pm0.5$ dex 
lower in \logg\ than our spectroscopic values.  Furthermore, the 20 objects in our 
sample classified as low-confidence WDs are spectroscopic 
$5\lesssim\log{g}\lesssim6$ low mass WDs.  Two objects in our sample that not listed 
in the \citet{fusillo21} catalog are the background halo star (J2338+1235) and a 
$P=15.28$ hr ELM WD binary, J0130$-$0530.

	\citet{pelisoli19b} perform a similar astrometric selection optimized for 
ELM WDs and identify 5762 candidates using Gaia DR2.  The 703 candidates with SDSS 
$17<g<19$ mag are drawn as green squares in Figure \ref{fig:plxcomp}.  The 
\citet{pelisoli19b} catalog includes half (31 of our 60) of our objects.  The 
matches are a 50/50 mix of ELM WDs and other WDs.  The other half of our objects are 
missing in the \citet{pelisoli19b} catalog, including the shortest-period ELM WD 
binaries.  This likely reflects their parallax selection and the Gaia DR2 parallax 
errors.

	While there is clear overlap between our ELM WD selection and existing 
Gaia-based catalogs, it is less clear how to compare the underlying selection 
criteria.  The \citet{fusillo21} catalog is focused on normal $\log{g}\sim8$ WDs, 
and so may provide a better means to exclude normal WDs (and possibly background 
halo stars, which should not exist in the WD catalog) from a search for ELM WDs than 
the other way around.  \citet{pelisoli19b} intentionally target ELM WDs but with low 
efficiency.  If we restrict the \citet{pelisoli19b} catalog to candidates with SDSS 
$17<g<19$ mag and $-0.4<(g-r)_0<-0.1$ mag -- our ELM WD sample should be essentially 
complete in this magnitude and color range -- then 4.5\% (16 of 357) of the 
\citet{pelisoli19b} candidates are ELM WDs with $5.5<\log{g}<7.2$.  Figure 
\ref{fig:plxcomp} suggests that the remaining candidates are likely WDs with higher 
$>$0.3 \msun\ masses. We conclude that the combination of multi-passband photometry 
and parallax makes for a more efficient selection of ELM WDs.

\subsection{New LISA Verification Binaries}

	One reason ELM WDs are interesting is that they reside in ultra-compact 
binaries.  Nearly two-thirds (35 of 57) of the candidates in the current sample are 
detached, double-degenerate binaries.  Two of the newly discovered binaries have 
orbital periods less than 32 minutes and are potential LISA gravitational wave 
sources.

	J1239$-$2041 was first identified as a 0.29 \msun\ ELM WD by 
\citet{kosakowski20} in ELM Survey South observations.  Follow-up motivated by the 
object's SDSS colors and Gaia parallax reveal it to be a $P = 22.514 \pm 0.186$ min 
binary.  The $K = 557.2 \pm 10.4$ \kms\ semi-amplitude requires a $M_2 > 0.61$ 
\msun\ minimum mass companion at $\simeq$0.25 \rsun\ separation.  This is a double 
degenerate binary, one that will merge in $<4$ Myr due to loss of energy and angular 
momentum via gravitational wave radiation.  Located at a heliocentric distance of 
$d=0.89\pm0.11$ kpc, the NASA LISA Detectability 
Calculator\footnote{\url{https://heasarc.gsfc.nasa.gov/lisa/lisatool/}} estimates 
LISA will detect J1239$-$2041 with a 4-yr SNR=12.

	J2322+2103 is a 0.25 \msun\ ELM WD in a $P=31.971 \pm 0.361$ min binary.  
Its $K = 248.1 \pm 4.3$ \kms\ semi-amplitude requires a $M_2 > 0.19$ \msun\ minimum 
mass companion with a $<30$ Myr gravitational wave merger time.  Located at a 
heliocentric distance of $d=1.04\pm0.16$ kpc, J2322+2103 has an estimated LISA 4-yr 
SNR=3.5.

\begin{figure} 	
 \includegraphics[width=3.5in]{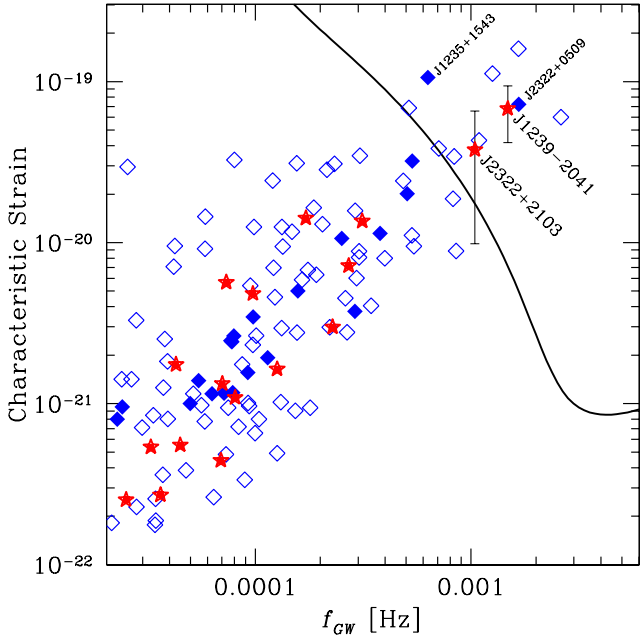}
 \caption{ \label{fig:gw}
	Characteristic strain versus gravitational wave frequency for the current 
sample (solid symbols) and all other ELM Survey binaries (open diamonds) compared to 
the LISA 4-yr sensitivity curve \citep{robson19}.  As before, new binaries are red 
and previously published binaries are blue. }
 \end{figure}

	Figure \ref{fig:gw} plots the distribution of characteristic strain versus 
gravitational wave frequency, $f_{GW}=2/P$, for the current sample (solid symbols) 
and all other ELM Survey binaries (open symbols) compared to the LISA 4-yr 
sensitivity curve \citep{robson19}.  New binaries are red, previously published 
binaries are blue.  The overall distribution reflects the fact that these detached 
WD binaries look like constant frequency sources to LISA, so their SNR is 
proportional to $\sqrt{N}$ orbits in 4 yr.  \citet{burdge20a} target similar sources 
with $P<1$ hr variability in the Zwicky Transient Factory and have discovered even 
shorter period systems \citep{burdge19, burdge20b}.

	Four of the 35 binaries in the current sample are likely LISA sources:  
J1239$-$2041 and J2322+2103, plus J1235+1543 \citep{kilic17} and J2322+0509
\citep{brown20b}.  All four are labeled in Figure \ref{fig:gw}.  Interestingly,
their LISA SNR may be underestimated.  Recent work shows that if the observed
metallicity-dependence of the stellar binary fraction is applied to WD models
\citep{thiele21}, or if the separation distribution of observed low mass WD binaries
is applied to all classes of WDs \citep{korol22}, then the predicted gravitational
wave foreground may be reduced by a factor of 2 for resolved LISA sources.  
Regardless, ELM WD binaries are a significant source of LISA ``verification
binaries,'' known binaries observable with both light and gravitational waves.

\section{Conclusions}

	We use Gaia parallax and SDSS multi-passband photometry to identify ELM WD 
candidates.  A joint parallax and color selection yields 57 ELM WD candidates with 
$17<g<19$ mag over the entire SDSS footprint in the color range $-0.4 < (g-r)_0 
< -0.1$ (approximately 9,000~K $<$ \teff\ $<$ 22,000~K).  Observations presented 
here complete the sample.

	The joint target selection is highly efficient for finding ELM WDs:  50\% of 
the candidates fall in the clean ELM WD region $5.5<\log{g}<7.2$ that overlaps the 
ELM WD evolutionary tracks.  All of the clean sample of ELM WDs are detached, 
double-degenerate binaries with $P<1$ day orbits.

	The 17 binaries published here bring the ELM Survey sample to 124 WD
binaries.  We find that 10\% of low mass WD binaries have $P<1$ hr orbits relevant
to LISA.  This implies that doubling the sample of low mass WDs, for example by
surveying the southern sky with SkyMapper as we have done with SDSS in the north,
should yield a dozen new multi-messenger WD binaries observable with LISA.

\begin{acknowledgements}

        We thank M. Alegria, B. Kunk, E.\ Martin, and A.\ Milone for their 
assistance with observations obtained at the MMT telescope, and Y.\ Beletsky for 
their assistance with observations obtained at Magellan telescope.  We thank 
the referee for constructive feedback that improved this paper. This research makes 
use the SAO/NASA Astrophysics Data System Bibliographic Service.  This work was 
supported in part by the Smithsonian Institution and in part by the NSF under grant 
AST-1906379.

The Apache Point Observatory 3.5-meter telescope is owned and operated by the 
Astrophysical Research Consortium.

Based on observations obtained at the Southern Astrophysical Research (SOAR) 
telescope, which is a joint project of the Minist\'{e}rio da Ci\^{e}ncia, Tecnologia 
e Inova\c{c}\~{o}es (MCTI/LNA) do Brasil, the US National Science Foundation's 
NOIRLab, the University of North Carolina at Chapel Hill (UNC), and Michigan State 
University (MSU).

Based on observations obtained at the international Gemini Observatory, a program of 
NSF's NOIRLab, which is managed by the Association of Universities for Research in 
Astronomy (AURA) under a cooperative agreement with the National Science Foundation. 
on behalf of the Gemini Observatory partnership: the National Science Foundation 
(United States), National Research Council (Canada), Agencia Nacional de 
Investigaci\'{o}n y Desarrollo (Chile), Ministerio de Ciencia, Tecnolog\'{i}a e 
Innovaci\'{o}n (Argentina), Minist\'{e}rio da Ci\^{e}ncia, Tecnologia, 
Inova\c{c}\~{o}es e Comunica\c{c}\~{o}es (Brazil), and Korea Astronomy and Space 
Science Institute (Republic of Korea).

\end{acknowledgements}

\facilities{MMT (Blue Channel spectrograph), Magellan:Clay (MagE spectrograph),
	SOAR (Goodman spectrograph), APO 3.5m (DIS spectrograph),
	Gemini:South (GMOS spectrograph)}

\software{IRAF \citep{tody86, tody93}, RVSAO \citep{kurtz98}, LCURVE \citep{copperwheat10}}


\end{document}